\documentclass[%
 reprint,
superscriptaddress,
 amsmath,amssymb,
 aps,
]{revtex4-1}

\usepackage[english]{babel}
\usepackage[caption=false]{subfig}
\usepackage{physics}
\usepackage{hyperref}
\usepackage{dblfloatfix}
\usepackage[dvipsnames]{xcolor} 
\usepackage{blindtext}
\usepackage{dblfloatfix}
\usepackage{graphicx}
\usepackage{dcolumn}
\usepackage{bm}
\usepackage{float}
\usepackage[pages=some,scale=1,angle=0,opacity=1]{background}

\begin{document}

\preprint{APS/123-QED}

\title{Dynamical behaviour of coupled atom-cavity systems in the single excitation limit}
\author{Ross Shillito}
\email{rshi896@aucklanduni.ac.nz}
\affiliation{Dodd-Walls Centre for Photonic \& Quantum Technologies, New Zealand}
\affiliation{Department of Physics, University of Auckland, Private Bag 92019, Auckland, New Zealand}
\author{Nikolett N\'emet}%
\email{nnem614@aucklanduni.ac.nz}
\affiliation{Dodd-Walls Centre for Photonic \& Quantum Technologies, New Zealand}
\affiliation{Department of Physics, University of Auckland, Private Bag 92019, Auckland, New Zealand}
\author{Scott Parkins}
\affiliation{Dodd-Walls Centre for Photonic \& Quantum Technologies, New Zealand}
\affiliation{Department of Physics, University of Auckland, Private Bag 92019, Auckland, New Zealand}

\begin{abstract}
We investigate the time evolution of the photon-detection probability at various output ports of an all-fiber coupled cavity-quantum-electrodynamics (cavity-QED) system. The setup consists of two atoms trapped separately in the field of two nanofiber cavities that are connected by a standard optical fiber. We find that the normal-mode picture captures well the main features of the dynamics. However, a more accurate description based on the diagonalization of a non-Hermitian Hamiltonian reveals the origin of small yet significant features in the spontaneous emission spectra.
\end{abstract}

\maketitle

\section{Introduction}
The successful implementation of quantum networks in a quantum computer holds the promise of revolutionising information processing and communication \cite{Zoller2005}. Due to their low propagation losses and scalability, all-fiber cavity-QED systems are excellent candidates for the construction of such a network \cite{KimbleH.J.2008TqiR, Solano2017, Nayak2018}. The cavities are formed from a short length of nanofiber sandwiched between a pair of fiber Bragg grating mirrors (FBGs), and an atom is trapped in the evanescent field of each nanofiber cavity \cite{Kato2015}. Atoms can act as qubits, forming the ‘nodes’ of a larger network with the cavities. Long, optical fibers which couple the cavities together operate as channels through which quantum information can be transferred.

\begin{widetext}
\begin{figure*}[b]
\vglue -.5cm
 \includegraphics[width=.9\textwidth]{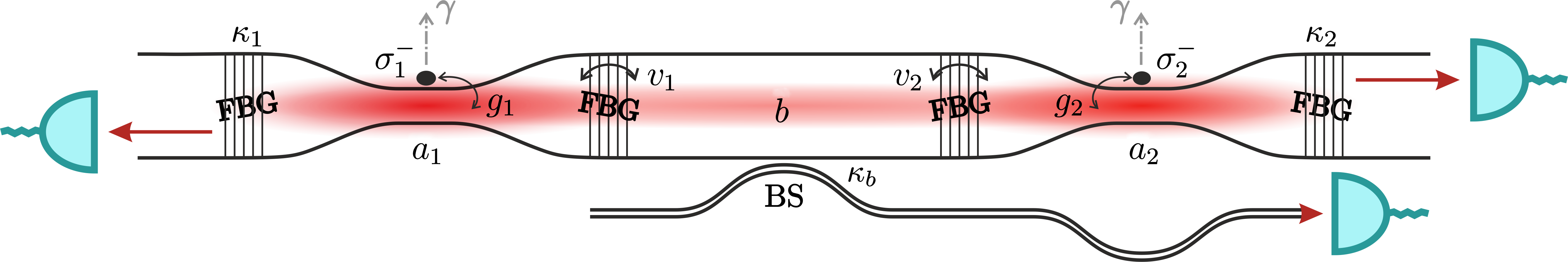}
 \caption{Schematic of fibre-coupled nanofiber cavity-QED systems, as realised in \cite{Kato2019,White2019}. The three optical cavities are constructed by four Fiber-Bragg grating mirrors (FBSs). The excitation can be detected through ports at Cavity 1, 2 or the connecting fiber via a beamsplitter (BS). The loss rates for cavity 1, 2 and the connecting fiber are given by $\kappa_1$, $\kappa_2$ and $\kappa_b$ respectively, and the atomic spontaneous emission rate by $\gamma$. The cavity-atom and cavity-fiber coupling strengths are $g_i$ and $v_i$, respectively. }
 \label{fig:ExperimentalSetup}
\end{figure*}
\end{widetext}

\begin{figure}
	\includegraphics[width=0.5\textwidth]{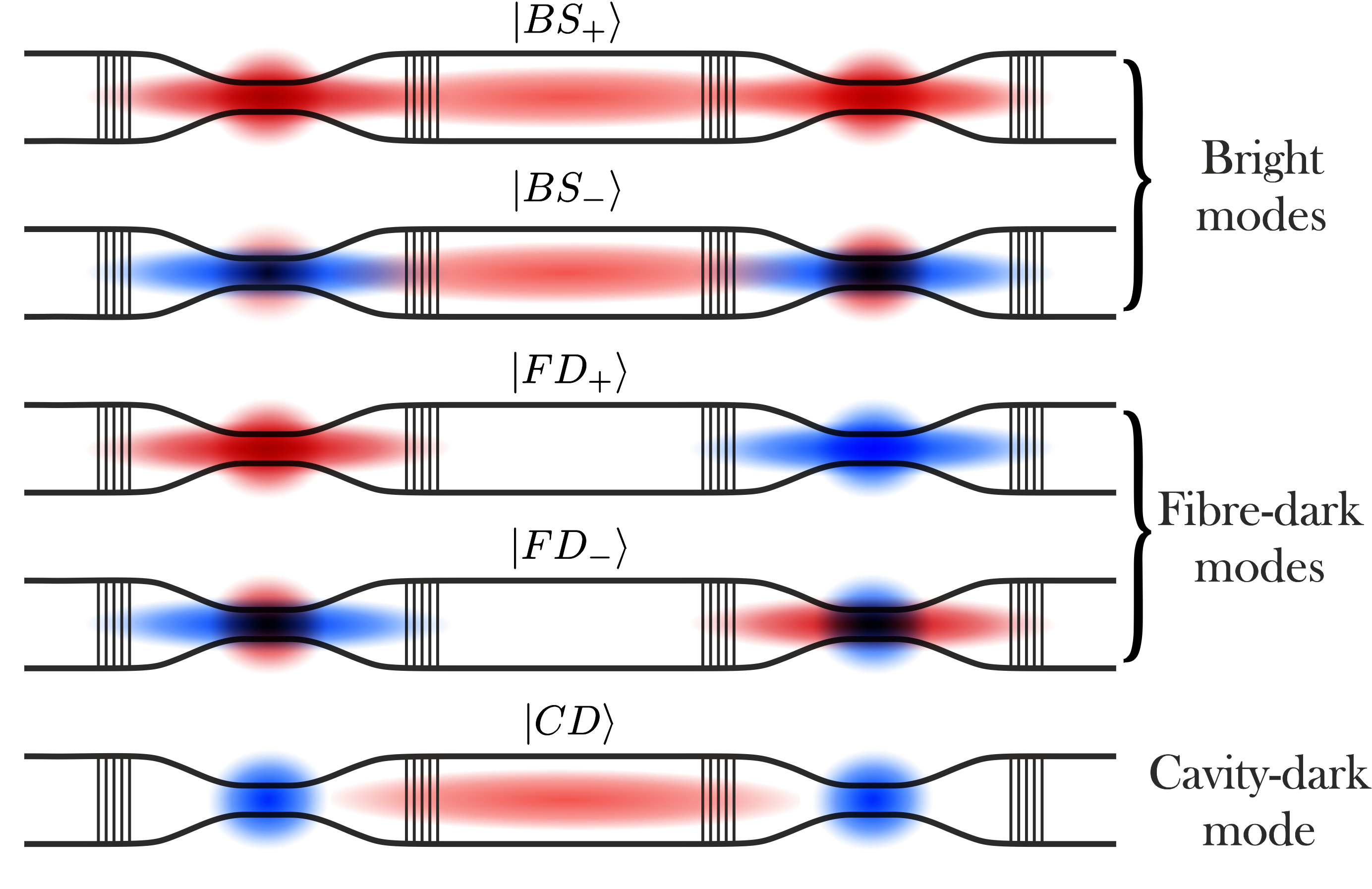}
		\vglue -.2cm
	\caption{A schematic of the five normal modes of the system. The shaded circles (ellipses) denote an atomic (field) excitation, and the colors denote their relative phases, so that a phase difference of $\pi$ is considered between the red and blue fields.}
	\label{fig:NormalModeSchematic}
\end{figure}

Such fiber-based, distributed systems also offer great potential for studying collective radiative properties of distantly separated atoms \cite{Solano20171, Sinha2017}. In this paper, we consider the theoretical model for an all-fiber coupled cavity-QED setup, as recently realised in \cite{Kato2019,White2019} and depicted in Fig. \ref{fig:ExperimentalSetup}, which -- in the considered single excitation limit -- can be pictured as five oscillators: two cavities, each containing an atom, and a length of fiber connecting these cavities together. If the connecting fiber is not too long, then it may be modelled  in terms of a single mode field. The dynamics of this system are best described in terms of the five normal modes, depicted in Fig. \ref{fig:NormalModeSchematic} -- two `bright states', $\ket{BS_\pm}$, in which all the optical modes are excited; two `fiber-dark' states, $\ket{FD_\pm}$, in which excitation in the connecting fiber is absent \cite{Kato2019}; and of most interest, the `cavity-dark' mode, $\ket{D}$, in which no excitation in either of the cavities is observed \cite{White2019}.

The experiments in  \cite{Kato2019,White2019} focused on the transmission spectrum of a weak probe laser and identified the various normal modes through distinct resonances in the spectrum. Here, we primarily focus instead on the explicit time evolution of system excitations, in particular as a result of an initial atomic excitation, in the absence of external driving. We explore how the various modes of the system respond to this excitation while varying the coupling strengths and decay rates. The observed dynamics undergoes a significant change due to these variations -- specifically, when one coupling strength significantly exceeds the other, contributions of certain normal modes can be suppressed or enhanced. 

With this in mind, arguably the richest and most complex dynamics occur when both the atomic and fiber coupling strengths are on the same order of magnitude. When decay is introduced into the system, the five normal modes have a finite bandwidth, and can thus overlap and interfere with one another. The degree of this interference is significant in the described parameter regime, and thus certain features of the cavity outputs cannot be explained using only non-interacting normal modes. Instead, we introduce new quasi-normal modes which arise from the diagonalization of a non-Hermitian Hamiltonian which, through an analytical and perturbative approach, well describe all of the features observed in the cavity outputs.

The paper is structured as follows: In Sec. \ref{sec:Theory}, we describe the open-system model, and how the evolution can be simplified to a system of five linear ordinary differential equations (ODE's). We consider this time evolution in the more natural normal-mode picture in Sec. \ref{sec:NormalModes}, then diagonalize the Hamiltonian in Sec. \ref{sec:Diagonalization} to yield analytical solutions for the amplitudes of the  fiber-dark modes $(A_\pm)$. We then explore the behaviour of the system in the strong coupling regime in Sec. \ref{sec:StrongCouplingRegime} and use perturbation theory to approximate the bright states and cavity-dark mode. Finally, we interpret the spectrum of spontaneous emission from both cavities in Sec. \ref{sec:SpectrumSpontaneousEmission}.

\section{Theoretical Model} \label{sec:Theory}
The Hamiltonian for this system is given by \cite{Kato2015}
\begin{equation} \label{Hi}
\begin{aligned}
	H &= \omega_c\left( a^\dagger_1 a_1 + a^\dagger_2 a_2 + b^\dagger b \right) +\sum_{i=1,2} \left( v_i^* a^\dagger_i b + v_i b^\dagger a_i\right) \\
	&+ \omega_a \left(\sigma_1^+ \sigma_1^- + \sigma_2^+ \sigma_2^-\right) + \sum_{i=1,2}\left(g_i a_i^\dagger \sigma_i^- +g_i^* \sigma_i^+ a_i \right),
\end{aligned}
\end{equation}
where $a_1, a_2, b$ are the annihilation operators for cavities 1 (on the left), 2 (on the right) and the connecting fiber, respectively, each degenerate with frequency $\omega_c$. The connecting fiber, referred to as `the fiber' for simplicity, is treated as a single mode. We investigate the case where the cavity and atomic transition frequencies are resonant, ($\omega_a=\omega_c$), and choose a frame rotating at this transition frequency. The parameters $v_i, g_i$ are the fiber-cavity and cavity-atom coupling strengths, respectively. Here we allow the coupling strengths to be complex -- however, without the presence of an external drive this phase dependence can be removed by a simple phase rotation of the operators. Thus, we can assume the coupling constants $v_i,g_i$ are positive and real, without loss of generality.

The master equation for this system is standard, and takes the form 
\begin{equation} \label{eq:MasterEquationFull}
\begin{aligned}
\dot{\rho} &= -i[H,\rho] + \kappa_1 \mathcal{D}[a_1]\rho +  \kappa_2 \mathcal{D}[a_2]\rho +   \kappa_b \mathcal{D}[b]\rho\\
&+ \frac{\gamma}{2} \left( \mathcal{D}[\sigma_1^-]\rho +  \mathcal{D}[\sigma_2^-]\rho \right).
\end{aligned}
\end{equation}
Each cavity has its own decay rate, $\kappa_i$, with a connecting fiber decay rate $\kappa_b$, which is significantly smaller than that of the cavities. We also consider vacuum spontaneous emission of the atomic excitation, with both atoms decaying at a fixed rate $\gamma/2\pi = 5.2$ MHz (cesium D2 line). 

In the single excitation limit, it is possible to decompose the density matrix into the sum of two parts: the ground state $\ket{G}$ and a one-quantum state $\ket{\psi(t)}$ \cite{Zeeb2015,Carmichael2007}:

\begin{equation}\label{eqn:densityispurish}
\rho(t) = P_G (t) \ketbra{G}{G} + \ketbra{\psi(t)},
\end{equation}
where $P_G(t)$ is the probability that a photon has been detected. The evolution of the state $\ket{\psi(t)}$ in the case of no photon detection is governed by the non-Hermitian Hamiltonian $\mathcal{H}$,
\begin{equation} \label{eq:NonHermitianHamiltonian}
\begin{aligned}
\mathcal{H} = H  -i& \left\lbrace \kappa_b b^\dagger b +  \sum_{i=1,2} \left( \kappa_i a^\dagger_i a_i +\frac{\gamma}{2} \sigma_i^+ \sigma_i^-\right)  \right\rbrace.
\end{aligned}
\end{equation}

The pure state $\ket{\psi(t)}$ can be written as the sum of the excited states, weighted by their corresponding probability amplitudes,

\begin{equation}
\ket{\psi(t)} = \sum_{i=1,2} \left[ \xi_i(t)\ket{A_i} + \alpha_i(t)\ket{C_i}\right] + \beta(t)\ket{B},
\end{equation}
where the excited states $\ket{A_i},\ket{C_i}, \ket{B}$ denote a single excitation in the corresponding atom, cavity and fiber mode, respectively. The equations of motion for the corresponding probability amplitudes are given by

\begin{equation} \label{eqns:CoupledProbAmplitudes}
\begin{aligned}
\dot{\xi}_1(t) &= -\frac{\gamma}{2} \xi_1(t) -i g_1\alpha_1(t),\\
\dot{\xi}_2(t) &= -\frac{\gamma}{2} \xi_2(t) -i g_2\alpha_2(t),\\
\dot{\alpha}_1(t) &= -\kappa_1\alpha_1(t) -i g_1\xi_1(t) -i v_1 \beta(t),\\
\dot{\alpha}_2(t) &= -\kappa_2\alpha_2(t) -i g_2\xi_2(t) -i v_2 \beta(t),\\
\dot{\beta}(t) &= -\kappa_b\beta(t)  -i v_1 \alpha_1(t) -i v_2 \alpha_2(t).
\end{aligned}
\end{equation}

\section{Normal modes} \label{sec:NormalModes}
Given that there are five coupled probability amplitudes in (\ref{eqns:CoupledProbAmplitudes}), it follows that the diagonalization of the Hamiltonian yields five normal modes. We focus on the  symmetric case, $(v_1=v_2=v)$, $(g_1=g_2=g)$, as this gives the greatest insight into the behaviour of the system. With this restriction, we obtain the following five states (Figure \ref{fig:NormalModeSchematic}):
\begin{equation} \label{eq:EnergyEigenstatesHamiltonian}
\begin{aligned}
\textrm{(i)}\  &\ket{BS_+}\propto g\ket{A_1} + g\ket{A_2} + \zeta \ket{C_1} \\
& + \zeta \ket{C_2} + 2v\ket{F}, &&\omega_0 + \zeta,\\
\textrm{(ii)}\  &\ket{BS_-}\propto g\ket{A_1} + g\ket{A_2} - \zeta \ket{C_1} \\
& - \zeta \ket{C_2} + 2v\ket{F}, &&\omega_0 - \zeta,\\
\textrm{(iii)}\  &\ket{FD_+} \propto \ket{A_1} - \ket{A_2} +\ket{C_1} -\ket{C_2}, \ \ \ &&\omega_0 +g, \\
\textrm{(iv)}\  &\ket{FD_-} \propto \ket{A_1} - \ket{A_2} - \ket{C_1} +\ket{C_2},\ \ \ &&\omega_0 -g, \\
\textrm{(v)}\  &\ket{CD} \propto -v\ket{A_1} -v\ket{A_2}+  g\ket{F},\ \ \ &&\omega_0.
\end{aligned}
\end{equation}
We observe two symmetric `bright states' $\ket{BS_\pm}$ , with normal mode splitting $\zeta = \sqrt{g^2 + 2v^2}$; two anti-symmetric `fiber-dark' modes $\ket{FD_\pm}$, with a splitting of $g$ and containing no contribution from the fiber; and a cavity-dark mode $\ket{CD}$, which has no contribution from either cavity \cite{White2019}.

It is possible to describe the evolution of the ket vector in terms of these normal mode amplitudes,
\begin{equation} \label{eq:NormalModeTransformPicture}
\ket{\psi_N(t)} = \sum_{\delta=+,-}\lbrace S_{\delta} \ket{BS_\delta} + A_{\delta}\ket{FD_\delta} \rbrace +D \ket{CD},
\end{equation}
where $S_{\pm}$, $A_{\pm}$ and $D$ refer to the probability amplitudes of the corresponding bright states, fiber-dark states and the cavity-dark state, respectively. These amplitudes can be described in terms of the original mode amplitudes,
\begin{equation} \label{eqs:TransformHamiltonianNormalModes}
\begin{aligned}
S_{\pm} &= \frac{1}{2\zeta}\left[g(\xi_1 + \xi_2) + 2v\beta \pm \zeta(\alpha_1 + \alpha_2) \right],\\
A_{\pm} &= \frac{1}{2}\left[(\xi_1 \pm \alpha_1) - (\xi_2 \pm \alpha_2) \right],\\
D &= \frac{1}{\zeta}\left[-v(\xi_1 + \xi_2) +g\beta \right],
\end{aligned}
\end{equation}
and we can re-express the original mode amplitudes in terms of their normal-mode counterparts,

\begin{equation} \label{eqns:opticalintermsofnormal}
\begin{aligned}
	\xi_{1,2} &=  \frac{1}{2\zeta}\left[g(S_+ + S_-) - 2vD \pm \zeta(A_+ + A_-)\right],\\
	\alpha_{1,2} &= \frac{1}{2}\left[(S_+ \pm A_+) - (S_- \pm A_-) \right],\\
	\beta &= \frac{1}{\zeta}\left[v(S_+ + S_-) +gD \right].
	\end{aligned}
\end{equation}

In case of a completely symmetric setup, 
the mode amplitudes $\left\lbrace S_\pm, D\right\rbrace$ and $\left\lbrace A_\pm \right\rbrace$  form a symmetric and anti-symmetric manifold respectively. They obey the equations of motion
\begin{equation} \label{eqns:MotionofNormalModesOriginal}
\begin{aligned}
\dot{S}_{\pm} &= -\left[\pm i\zeta + \frac{\Gamma_{S+}}{2}\right]S_{\pm} - \frac{\Gamma_{S-}}{2}S_{\mp} +\Gamma_{SD}D,\\
\dot{D} &=-\Gamma_DD +\Gamma_{SD}(S_+ + S_-),\\
\dot{A}_{\pm} &= -\left[\pm ig + \frac{\Gamma_{A+}}{2}\right]A_{\pm} - \frac{\Gamma_{A-}}{2}A_{\mp},
\end{aligned}
\end{equation}
where we introduce new decay rates
\begin{equation} \label{eqn:NormalModeDecayRates}
\begin{aligned}
\Gamma_{S\pm} &= \frac{g^2 \gamma/2 + 2v^2 \kappa_b}{\zeta^2} \pm \kappa, \quad &&\Gamma_{A\pm} = \frac{\gamma}{2} \pm \kappa,\\
\Gamma_{SD} &= \left( \frac{\gamma}{2} - \kappa_b \right) \frac{vg}{\zeta^2},\quad  &&\Gamma_{D} = \frac{\gamma v^2 + g^2\kappa_b}{\zeta^2}.
\end{aligned}
\end{equation}

This means that we can independently solve for the anti-symmetric and symmetric contributions to the ket vector,

\begin{equation}
	\ket{\psi(t)} = \ket{\psi_A(t)} +\ket{\psi_S(t)}, 
\end{equation}
by solving (\ref{eqns:MotionofNormalModesOriginal}) with the appropriate initial conditions. Without loss of generality, we choose Atom 1 to begin in the excited state for the entirety of this paper. This gives us the following initial conditions for the normal modes:

\begin{equation} \label{eqn:NormalModeIC}
S_{\pm}(0) = \frac{g}{2\zeta} ,\quad A_{\pm}(0) = \frac{1}{2}, \quad D(0)  = -\frac{v}{\zeta}.
\end{equation}

\section{Diagonalization} \label{sec:Diagonalization}
In order to quantify the behaviour of the optical and normal modes, it is informative to diagonalize the non-Hermitian Hamiltonian $\mathcal{H}$ in the normal mode basis. Let $V$ be the transformation matrix such that 
\begin{equation}
\label{eq:diag}
-i\mathcal{H} = VDV^{-1}.
\end{equation}
Then the rows of $V^{-1}$ and columns of $V$ are the left and right eigenvectors of this system, respectively. We shall refer to the right eigenvectors as the \textit{quasi-normal modes}, labelled as $\ket{QBS_{\pm}},\ket{QFD_{\pm}},\ket{QCD}$, given their close association with the normal modes.  These states are trivial to propagate,
\begin{equation}\label{eqn:PropQuasiStates}
-i\mathcal{H}\ket{\mu_i} = \lambda_i \ket{\mu_i},
\end{equation}
where $\ket{\mu_i}$ is an arbitrary right eigenvector. Thus, without loss of generality, we can describe the evolution of the system with Atom 1 initially excited as
\begin{equation} \label{eqn:EvolveasQuasi}
\ket{\psi(t)} =  \sum_{i=1}^{5} w_{i1} e^{\lambda_i t}\ket{\mu_i},
\end{equation}
where $w_{i1}$ are the weights of the contributions from Atom 1 to the left eigenvectors $\bra{\varphi_i}$. Transforming back into the normal mode basis via operator $V$, with matrix elements $v_{ij}$, yields the solutions to the probability amplitudes $\left.d_i \in \left\lbrace A_\pm, S_\pm, D \right\rbrace \right.$:
\begin{equation}
d_i(t) = \sum_{j=1}^{3}\Lambda_{ij}e^{\lambda_j t}, \quad  \Lambda_{ij} = v_{ij}w_{j1}.
\label{eq:Eqnswiththreecoeffs}
\end{equation}
Given that the two manifolds can be solved independently, these coefficients will oscillate with at most three eigenvalues. We can finally use (\ref{eqns:opticalintermsofnormal}) to solve for the original mode amplitudes ($c_i \in \{\xi_{1,2},\alpha_{1,2},\beta\}$), yielding a similar result to (\ref{eq:Eqnswiththreecoeffs}), albeit with up to five eigenvalues and transformed coefficients $\chi_{ij}$:

\begin{equation}
c_i(t) = \sum_{j=1}^{5}\chi_{ij}e^{\lambda_j t}. 
\label{eq:Eqnswithfivecoeffs}
\end{equation}

\subsection{Anti-symmetric manifold}

We begin our analysis with the simpler, anti-symmetric manifold, equivalent to solving the matrix-vector equation
\begin{equation} \label{eqn:Matrixvec}
	\frac{d}{dt} \begin{bmatrix}
	A_+\\A_-
	\end{bmatrix} =
	\begin{bmatrix}
	-ig - \Gamma_{A+}/2 &   - \Gamma_{A-}/2\\
	 - \Gamma_{A-}/2 & ig - \Gamma_{A+}/2 &
	\end{bmatrix}
	\begin{bmatrix}
	A_+\\A_-
	\end{bmatrix}.
\end{equation}
Diagonalizing the matrix in (\ref{eqn:Matrixvec}) trivially yields the desired matrices $V$ and $D$ in equation (\ref{eq:diag}). We, thus, find the quasi-normal modes and associated eigenvalues to be of the form

\begin{equation} \label{eqn:QuasiRightFDandevalue}
\begin{aligned}
	\ket{QFD_{\pm}} &= \frac{2i(g\pm p)}{\Gamma_{A-}}\ket{FD_+} + \ket{FD_-},\\
	 \lambda_{A\pm} &= -\frac{\Gamma_{A+}}{2} \pm ip,
	\end{aligned}
\end{equation}
where $p=\sqrt{g^2 - \left(\frac{\Gamma_{A-}}{2}\right)^2}$ and $\left\lbrace p,\  \Gamma_{A-} \right\rbrace \neq 0$. We intentionally leave this state unnormalized, as any normalization constant would only amount to scaling the matrix inverse $V^{-1}$. The left eigenvectors  $\bra{QFD_\pm}$ are obtained from this matrix inverse,

\begin{equation} \label{eqn:LeftFDandevalue}
	\bra{QFD_\pm} = \mp\frac{i\Gamma_{A-}}{4p} \bra{FD_+} + \frac{(p \mp g)}{2p} \bra{FD_-}.
\end{equation}

While the left and right eigenvectors have different forms, it should be no surprise that

\begin{equation}
\bra{\phi_i}\ket{\mu_j} = \delta_{ij},
\end{equation}
which is guaranteed by the condition $V V^{-1} = \mathbb{I}_2$. 

Using (\ref{eqn:NormalModeIC}), (\ref{eqn:PropQuasiStates}) and (\ref{eqn:LeftFDandevalue}) we can obtain the coefficients $w_{i1}$, and thus write the anti-symmetric contribution to our state in terms of the quasi-normal modes,

\begin{equation}
\begin{aligned}
\ket{\psi_A(t)} = &-\frac{i\Gamma_{A-} + 2(g-p)}{8p}e^{\lambda_+t}\ket{QFD_+}\\
&+ \frac{i\Gamma_{A-} + 2(g+p)}{8p}e^{\lambda_-t}\ket{QFD_-}.
\end{aligned}
\end{equation}

Finally, we use (\ref{eqn:QuasiRightFDandevalue}) to obtain an analytical solution for the fiber-dark probability amplitudes,

\begin{equation}\label{eqn:AnalyticalFiberDark}
\hspace{-1.3cm}
\begin{aligned}
	A_\pm(t) = \frac{e^{-\frac{\Gamma_{A+} t}{2}}}{4ip}
	&\left(e^{-ipt}\left[i(p\pm g)+\frac{\Gamma_{A-}}{2}\right] \right.\\
	 &+ \left.e^{ipt}\left[i(p \mp g)-\frac{\Gamma_{A-}}{2}\right] \right).
	\end{aligned}
\end{equation}

A few things are immediately obvious from the above solution:

\begin{itemize}
	\item $A_+(t) = A^*_-(t)$;
	\item The amplitudes decay as $\frac{\Gamma_{A+}}{2}$;
	\item When $p$ is real, 
	\begin{itemize}
		\item the fiber-dark modes are \textit{underdamped}, and experience oscillations;
	\item	the amplitudes $A_\pm$ predominately oscillate as $\pm p$;
	\item as $\frac{\Gamma_{A-}}{2}$ grows, the oscillations gain a stronger contribution from the corresponding value $\mp p$;
		\item Oscillatory behavior ceases at the critical values $\frac{\Gamma_{A-}}{2} = \pm g$.
	\end{itemize} 	
\item When $p$ is imaginary, we say the system is \textit{overdamped}, and decays as a sum of two appropriately weighted exponentials.
\end{itemize}

When $p=0$, we say the system is \textit{critically damped}; the equations simplify to

\begin{equation}
	A_\pm(t) = \frac{e^{-\frac{\Gamma_{A+} t}{2}}}{2} \left[1 - \frac{\Gamma_{A-} t}{2}(1\pm i)\right].
\end{equation}

\subsection{Symmetric manifold}

While an analytical solution for the symmetric manifold exists, it is not needed to understand the properties of this system -- indeed, in the strong coupling limit when $g \gg v$ or $v \gg g$, $\Gamma_{SD} \rightarrow 0$, we find the symmetric manifold decouples the cavity-dark mode (see Sec. \ref{sec:AtomDomCoupling} and \ref{sec:Fiberdominatedcoupling}). When the coupling strengths are on the same order of magnitude, we can take a perturbative approach to approximate the manifold, as done in Sec. \ref{sec:Pertapproach}.

\section{Strong coupling regime} \label{sec:StrongCouplingRegime}
In the strong coupling regime the cavity-atom and/or fiber-cavity coupling strengths exceed the decay rates of the system $(g,v > \kappa, \frac{\gamma}{2})$. In the cases $v\gg g$ and $g \gg v$, we are able to simplify the equations of motion presented in (\ref{eqns:MotionofNormalModesOriginal}), and use (\ref{eqn:AnalyticalFiberDark}) to obtain an expression for the behaviour of the normal modes, cavities, and atoms. When the two coupling strengths are comparable, such an elegant solution does not exist -- instead, we can approximate the solution by taking a perturbative approach. 

\subsection{Atom dominated coupling}
\label{sec:AtomDomCoupling}
\begin{figure}[t]
	\includegraphics[]{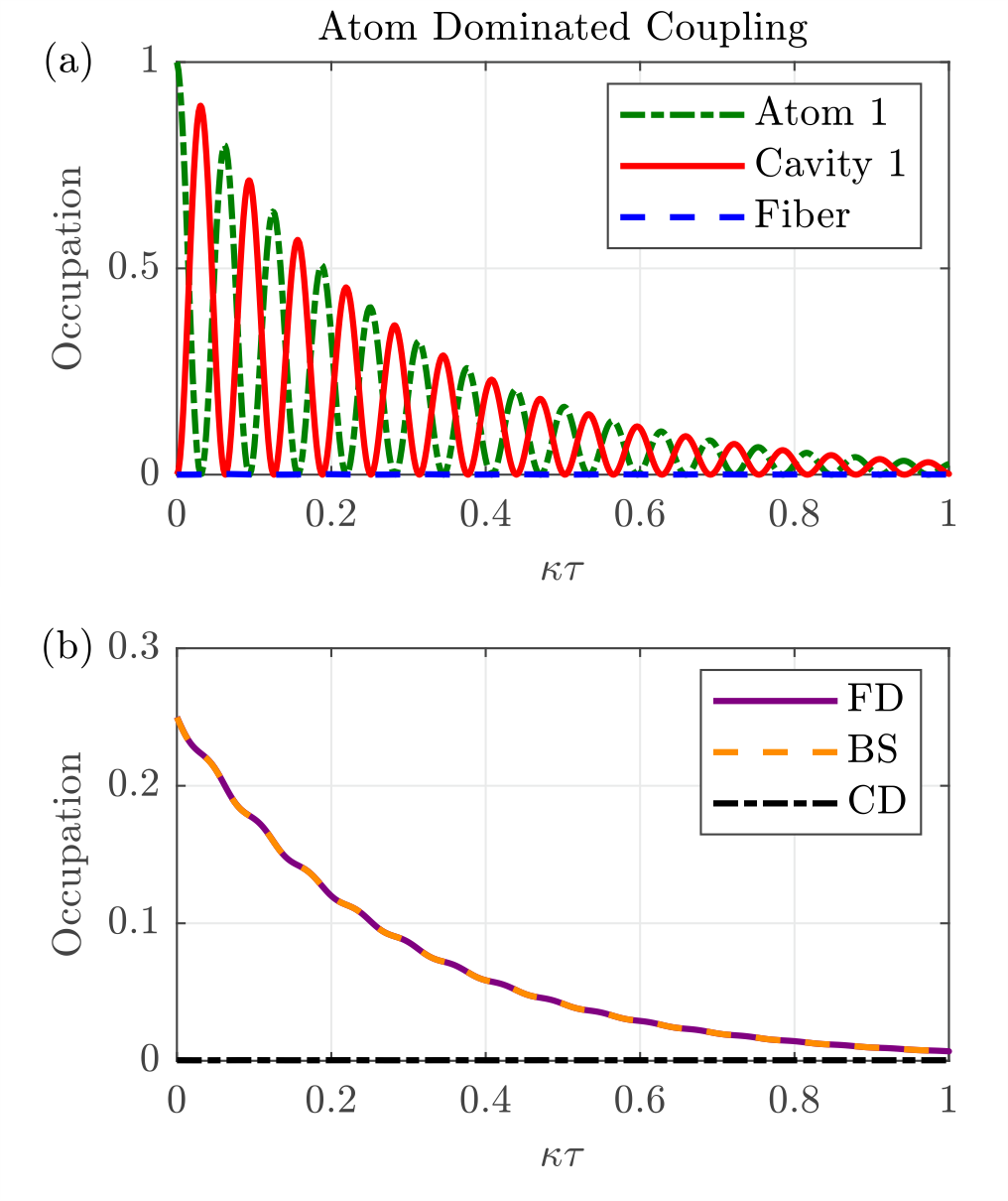}
	\vglue -.5cm
	\caption{Relative excitations of the (a) optical and (b) normal modes in the atom-dominated coupling regime. $[\kappa_b,\kappa,v,g]=[0.01,1,1,50]$ (units of $2\pi \cdot$MHz). Atom 1 is initially excited. }
	\label{fig:AtomDomOscillations}
\end{figure}

When $g \gg v$, the coupling between the cavities and the fiber is almost negligible, ensuring that the excitation is essentially confined to Atom 1 and Cavity 1, as can be seen in Figure \ref{fig:AtomDomOscillations}(a). In this regime the normal mode splitting
$\zeta \rightarrow g$, giving us the initial conditions for the normal mode amplitudes as
\begin{equation}
S_{\pm}(0) = \frac{1}{2} ,\quad A_{\pm}(0) = \frac{1}{2}, \quad D(0)  = 0.
\end{equation}

We also find that (\ref{eqns:MotionofNormalModesOriginal}) further decouples, as $\Gamma_{SD} \rightarrow 0$. Moreover, we find $\Gamma_{S\pm} \rightarrow \Gamma_{A\pm}$, resulting in the equations of motion 
\begin{equation} \label{eqns:MotionofNormalModesAtomDom}
\begin{aligned}
\dot{S}_{\pm} &= -\left[\pm i g + \frac{\Gamma_{A+}}{2}\right]S_{\pm} - \frac{\Gamma_{A-}}{2}S_{\mp} ,\\
\dot{A}_{\pm} &= -\left[\pm ig + \frac{\Gamma_{A+}}{2}\right]A_{\pm} - \frac{\Gamma_{A-}}{2}A_{\mp}.
\end{aligned}
\end{equation}
Thus, with the same initial conditions for the bright state and fiber-dark state amplitudes, $S_\pm, A_\pm$, it is trivial to see that $S_\pm = A_\pm$. We also note small oscillations in the normal mode occupations, as seen in Figure \ref{fig:AtomDomOscillations} (b) -- these are due to the interactions within each set of manifolds $\left\lbrace S_\pm\right\rbrace$ and $\left\lbrace A_\pm \right\rbrace$, which scale with $\Gamma_{A-}$. These oscillations could thus be eliminated by setting the cavity decay rate $\kappa = \frac{\gamma}{2}$. The cavity-dark mode is not excited in this regime.

Note that, using (\ref{eqns:opticalintermsofnormal}) and (\ref{eqns:MotionofNormalModesAtomDom}), the solutions for the atomic and cavity probability amplitudes are readily obtained as:

\begin{equation}
\begin{aligned}
	\xi_1(t) &= e^{-\frac{\Gamma_{A+} t}{2}}\left[\cos(pt) + \frac{\Gamma_{A-}}{2p}\sin(pt) \right],\\
	\alpha_1(t) &= \frac{-ig}{p}e^{-\frac{\Gamma_{A+} t}{2}}\sin(pt).
	\end{aligned}
\end{equation}

\subsection{Fiber dominated coupling}
\label{sec:Fiberdominatedcoupling}
\begin{figure}[]
	\includegraphics[width =\columnwidth]{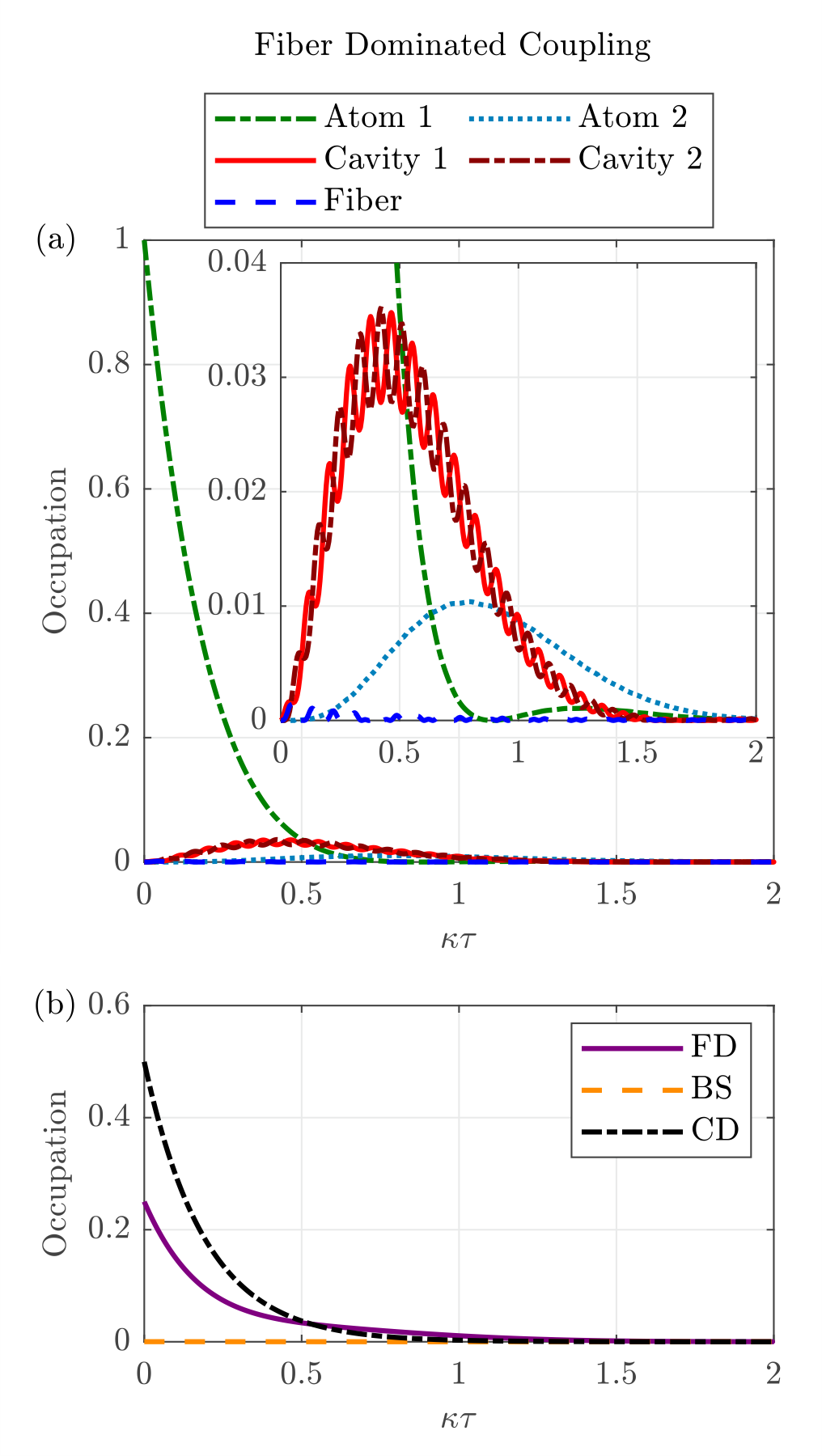}
		\vglue -.5cm
	\caption{Relative excitations of the (a) optical and (b) normal modes in the fiber-dominated coupling regime. The inset in (a) shows the curves on smaller horizontal and vertical scales.  The bright states are not excited. $[\kappa_b,\kappa,v,g]=[0.01,1,50,2]$. Atom 1 is initially excited. }
	\label{fig:FiberDomOscillations}
\end{figure}

When $v \gg g$, we once again find the cavity-dark mode decoupling from the bright states, as $\Gamma_{SD} \rightarrow 0$. In this regime the normal mode splitting $2\zeta \rightarrow 2\sqrt{2}v$, giving us the following initial conditions for the normal mode amplitudes
\begin{equation}
S_{\pm}(0) \approx 0 ,\quad A_{\pm}(0) = \frac{1}{2}, \quad D(0)  = \frac{-1}{\sqrt{2}}.
\end{equation}
The relevant normal mode amplitudes obey the equations of motion 
\begin{equation} \label{eqns:MotionofNormalModesFiberDom}
\begin{aligned}
\dot{A}_{\pm} &= -\left[\pm i g + \frac{\Gamma_{A+}}{2}\right]A_{\pm} - \frac{\Gamma_{A-}}{2}A_{\mp},\\
\dot{D} &=-\frac{\gamma}{2} D.
\end{aligned}
\end{equation}
Note that the occupation of the bright states is almost negligible, but produces rapid fluctuations in the occupations of Cavity 1 and 2. In the original picture, we observe an approximately exponential decay in the atomic excitation, with negligible occupation in the fiber mode, as can be seen in Figure \ref{fig:FiberDomOscillations}.

In the limit $v\rightarrow \infty$, we can once again obtain an analytical solution for the cavity and atomic probability amplitudes. Assuming $p\neq 0$, which is true in the strong coupling regime (as $g > \kappa, \gamma/2$), we find

\begin{equation}
\label{eqns:OpticalAmplitudesAnswerFiber}
\begin{aligned}
	\alpha_{1}(t) &= \frac{ig}{p}e^{-\frac{\Gamma_{A+} t}{2}}\sin(pt),\\
	\alpha_{2}(t) &= -\alpha_1(t),\\
\xi_{1}(t) &= \frac{1}{2}e^{-\frac{\gamma t}{2}} + e^{-\frac{\Gamma_{A+} t}{2}}\left[\cos(pt) + \frac{\Gamma_{A-}}{2p}\sin(pt) \right],\\
\xi_{2}(t) &= \frac{1}{2}e^{-\frac{\gamma t}{2}} - e^{-\frac{\Gamma_{A+} t}{2}}\left[\cos(pt) + \frac{\Gamma_{A-}}{2p}\sin(pt) \right].
\end{aligned}
\end{equation}
Note that these equations still hold in the weak coupling regime, where $g < \kappa, \gamma/2$. In the critically damped case where $p=0$, we find that the oscillations in the optical and atomic mode amplitudes cease, and the equations simplify to

\begin{equation}
\begin{aligned}
\alpha_{1}(t) &= \frac{igt}{2}e^{-\frac{\Gamma_{A+}t}{2}},\\
\alpha_{2}(t) &= -\alpha_1(t),\\
\xi_{1}(t) &= \frac{1}{2}e^{-\frac{\gamma t}{2}} + \frac{1}{2}e^{-\frac{\Gamma_{A+} t}{2}}\left(1-\frac{\Gamma_{A-}t}{2}\right),\\
\xi_{2}(t) &= \frac{1}{2}e^{-\frac{\gamma t}{2}} - \frac{1}{2}e^{-\frac{\Gamma_{A+} t}{2}}\left(1-\frac{\Gamma_{A-}t}{2}\right).
\end{aligned}
\end{equation}

We find that (\ref{eqns:OpticalAmplitudesAnswerFiber}) still holds in the over-damped case, where $p$ becomes imaginary -- this amounts to a \textit{slower} decay in the amplitudes, and no oscillations are observed.

\subsection{Perturbative approach to Comparable Coupling} \label{sec:Pertapproach}
When the two coupling strengths $v$ and $g$ are comparable, all the optical and normal modes are excited -- in particular, we cannot eliminate the fiber-mode. Investigating the time evolution of the average cavity occupations using quantum trajectories, the role of the fiber mode becomes clear as it mitigates the excitation exchange between the two cavities (Figure \ref{fig:Comparablescillations}). In order to describe the evolution analytically the best approach is through perturbation theory which, in the strong coupling regime, proves to be an excellent approximation. 

We begin our approach by assuming that $\kappa_b \approx 0$, which simplifies the coupling decay rates in (\ref{eqn:NormalModeDecayRates}) considerably to

\begin{equation} \label{eqn:simplifiedNormalModeDecayRates}
\begin{aligned}
\Gamma_{S\pm} &\approx \frac{g^2\gamma}{2\zeta^2} \pm \kappa,\\
\Gamma_{SD} &\approx \frac{\gamma vg}{2\zeta^2},\\
\Gamma_{D} &\approx \frac{\gamma v^2}{\zeta^2}.
\end{aligned}
\end{equation}
For simplicity, we assume $\Gamma_{S-} \approx 0$ -- a condition which is satisfied as long as $\kappa \approx \gamma/6$, where $v \approx g$. This reduces the matrix equation for the symmetric manifold to
\begin{equation}
\frac{d}{dt}
	\begin{bmatrix}
	S_+ \\
	 S_- \\
	 D
	\end{bmatrix}
	\approx
	\begin{bmatrix}
	-\frac{\Gamma_S+}{2} - i\zeta & 0 & \Gamma_{SD}\\
	0 & -\frac{\Gamma_S+}{2} + i\zeta & \Gamma_{SD}\\
	\Gamma_{SD} & \Gamma_{SD} & \Gamma_D\\
	\end{bmatrix}
		\begin{bmatrix}
	S_+ \\
	S_- \\
	D
	\end{bmatrix}.
\end{equation}

We seek to treat the off-diagonal elements $\Gamma_{SD}$ as a perturbation $P$ to the system, leaving the unperturbed Hamiltonian $\mathcal{H}_0$ as a diagonal matrix:

\begin{equation}
	\begin{aligned}
	\mathcal{H}_0 &=
	\begin{bmatrix}
	-\frac{\Gamma_S+}{2} - i\zeta & 0 & 0\\
	0 & -\frac{\Gamma_S+}{2} + i\zeta & 0\\
	0 & 0 & \Gamma_D\\
	\end{bmatrix}
	,\\
	 P &= 
	\begin{bmatrix}
0 & 0 & \Gamma_{SD}\\
0 & 0 & \Gamma_{SD}\\
\Gamma_{SD} & \Gamma_{SD} & 0
\end{bmatrix}.
	\end{aligned}
\end{equation}

 In the strong coupling regime $\zeta \gg \Gamma_{SD}$, meaning that the perturbation is significantly smaller than the diagonal elements for $S_+$ and  $S_-$. The ratio $\Gamma_{D}/\Gamma_{SD} \approx 2v/g$ ensures that this perturbation is most accurate when $v > g/2$ (please see the Appendix for more details regarding this approximation). 

The zeroth order quasi-normal modes and eigenvalues are trivially obtained from $\mathcal{H}_0$,

\begin{equation}
\begin{aligned}
	\ket{QBS_{\pm}}_0 &=\ket{BS_\pm},\\
	\ket{QCD}_0 &= \ket{CD},\\
	 \lambda_{S\pm} &= -\left(\frac{\Gamma_{S+}}{2}\pm i\zeta\right),\\
	 \lambda_{D} &= -\Gamma_D.
	\end{aligned}
\end{equation}
Our approach to the perturbation is standard:
\begin{equation}
\begin{aligned}
\Delta\lambda &= {}_{0}\bra{\mu_i}P\ket{\mu_i}_0,\\
\ket{\mu_i}_1 &= \ket{\mu_i}_0 +\sum_{\substack{j=1\\ j\neq i}}^{N} \frac{{}_{0}\bra{\mu_j}P\ket{\mu_i}_0}{\lambda_{0i} - \lambda_{0j}} \ket{\mu_j}_0.
\end{aligned}
\end{equation}
Using this approach, we quickly find

\begin{equation}\label{eqns:pertrightstates}
\begin{aligned}
	\ket{QCD}_1 &= \ket{CD}  -\Delta_{S+}\ket{BS_+}-\Delta_{S-}\ket{BS_-},\\
	\ket{QBS_{\pm}}_1 &= \ket{BS_{\pm}}  +\Delta_{S\pm}\ket{CD},\\
	\Delta\lambda_1 &=0 \  \textrm{for all } \lambda,
\end{aligned}
\end{equation}
where
\begin{equation}
\begin{aligned}
\Delta_{S\pm} = \frac{\Gamma_{SD}}{\Gamma_D -\frac{\Gamma_{S+}}{2} \mp i\zeta}.\\
\end{aligned}
\end{equation}
A similar perturbative approach can be taken to find the left eigenstates, which are of the same form:

\begin{equation} \label{eqns:pertleftstates}
\begin{aligned}
	\bra{QCD}_1 &= \bra{CD}  -\Delta_{S+}\bra{BS_+} -\Delta_{S-}\bra{BS_-},\\
	\bra{QBS_{\pm}}_1 &= \bra{BS_{\pm}}  +\Delta_{S\pm}\bra{CD}.
\end{aligned}
\end{equation}

Using (\ref{eqn:EvolveasQuasi}) with the initial conditions given in (\ref{eqn:NormalModeIC}) and the matrix inverse $V^{-1}$ constructed from the left eigenvectors in (\ref{eqns:pertleftstates}), we can write the symmetric contribution to the ket vector ($\ket{\psi(t)}$ in (\ref{eqn:densityispurish})), written in terms of the quasi-normal modes:

\begin{figure}[t]
		\includegraphics[]{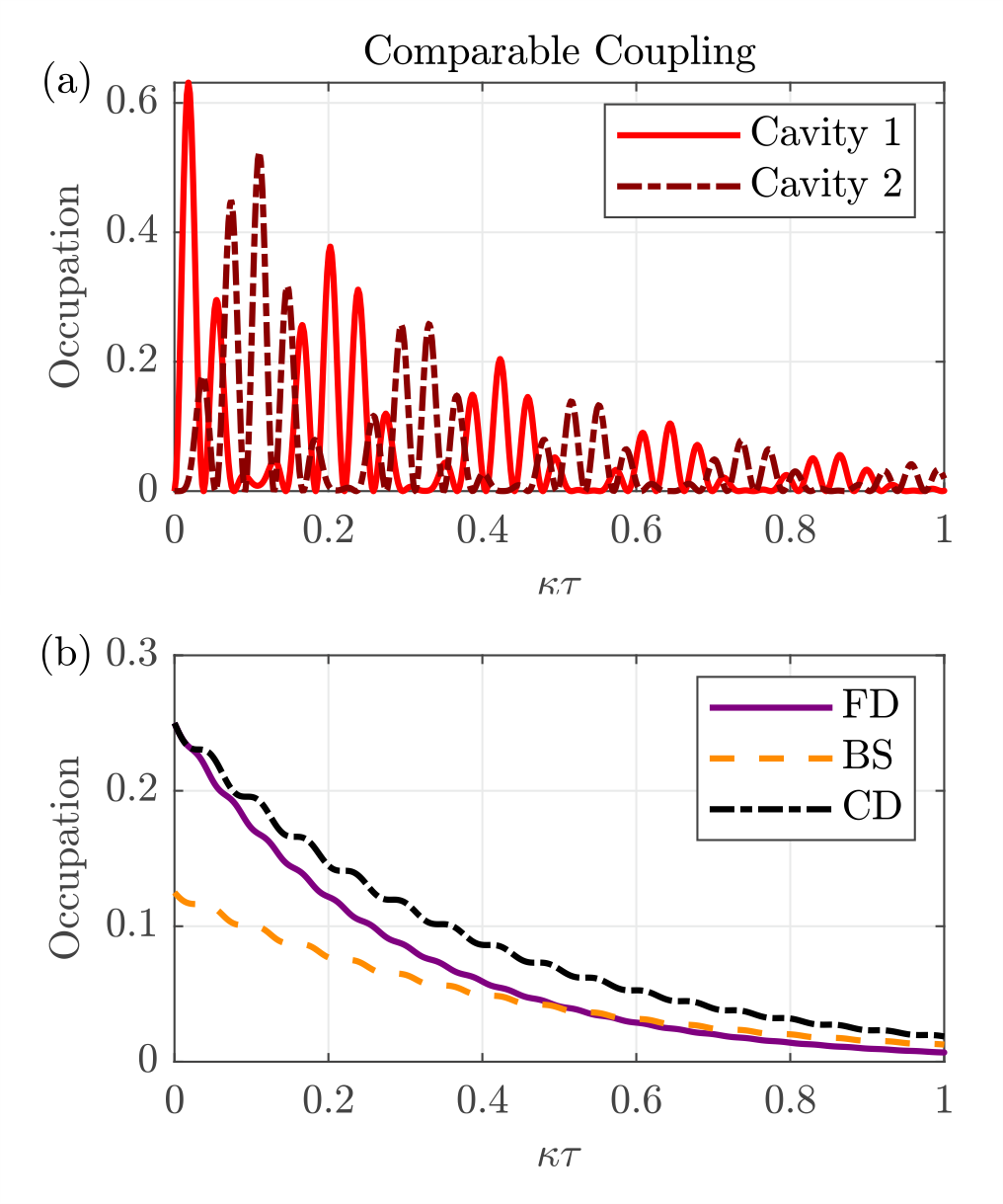}
\vglue -.5cm
	\caption{Relative excitations of the (a) optical and (b) normal modes when the coupling strengths are comparable. $[\kappa_b,\kappa,v,g]=[0.01,1,50,50\sqrt{2}]$ (units of $2\pi \cdot$MHz). Atom 1 is initially excited.}
	\label{fig:Comparablescillations}
\end{figure}

\begin{equation}
\begin{aligned}
	\ket{\psi_S(t)} \approx f_+(t)\ket{QBS_+} + f_-(t)\ket{QBS_+} + g(t)\ket{QCD}),
	\end{aligned}
\end{equation}
where
\begin{equation} \label{eqns:QuasiBrightStateTimePerturbed}
\begin{aligned}
f_\pm(t) &= \frac{1}{\zeta}\left[ \left(\frac{g}{2}-v\Delta_{S\pm}\right)e^{-\left(\frac{\Gamma_{S+}}{2} \pm i\zeta \right)t}\right],\\
g(t) &= \frac{1}{\zeta}\left[ \left(-\frac{g}{2}\Delta_{S+} - \frac{g}{2}\Delta_{S-} -v\right)e^{-\Gamma_D t}\right].
\end{aligned}
\end{equation}
We can then revert back to the normal mode basis using (\ref{eqns:pertrightstates}) and obtain, to a reasonable approximation, the equations for the amplitudes in the symmetric manifold:

\begin{equation}
\begin{aligned}
S_{\pm}(t) &= f_{\pm}(t) -\Delta_{S\pm} g(t),\\
D(t) &= g(t) + \Delta_{S+}f_+(t) + \Delta_{S-}f_-(t),
\end{aligned}
\end{equation}
and thus, the equations for the probability amplitudes of the two cavities,
\begin{equation}\hspace{-0.3cm} \label{eqn:ApproxtoCavitiesTime}
\begin{aligned}
\alpha_{1}(t) &= \frac{1}{2}\left[(f_+(t) -f_-(t)) + (A_+(t) - A_-(t)) \right]\\
 &- \frac{1}{2}(\Delta_{S+} - \Delta_{S-})g(t),\\
 \alpha_{2}(t) &= \frac{1}{2}\left[(f_+(t) -f_-(t)) -(A_+(t) - A_-(t))  \right]\\
 &-  \frac{1}{2}(\Delta_{S+} - \Delta_{S-})g(t).
\end{aligned}
\end{equation}
where $A_\pm(t)$ are defined in (\ref{eqn:AnalyticalFiberDark}). 

\begin{widetext}
\begin{figure*}[t]
	\includegraphics[width =2\columnwidth]{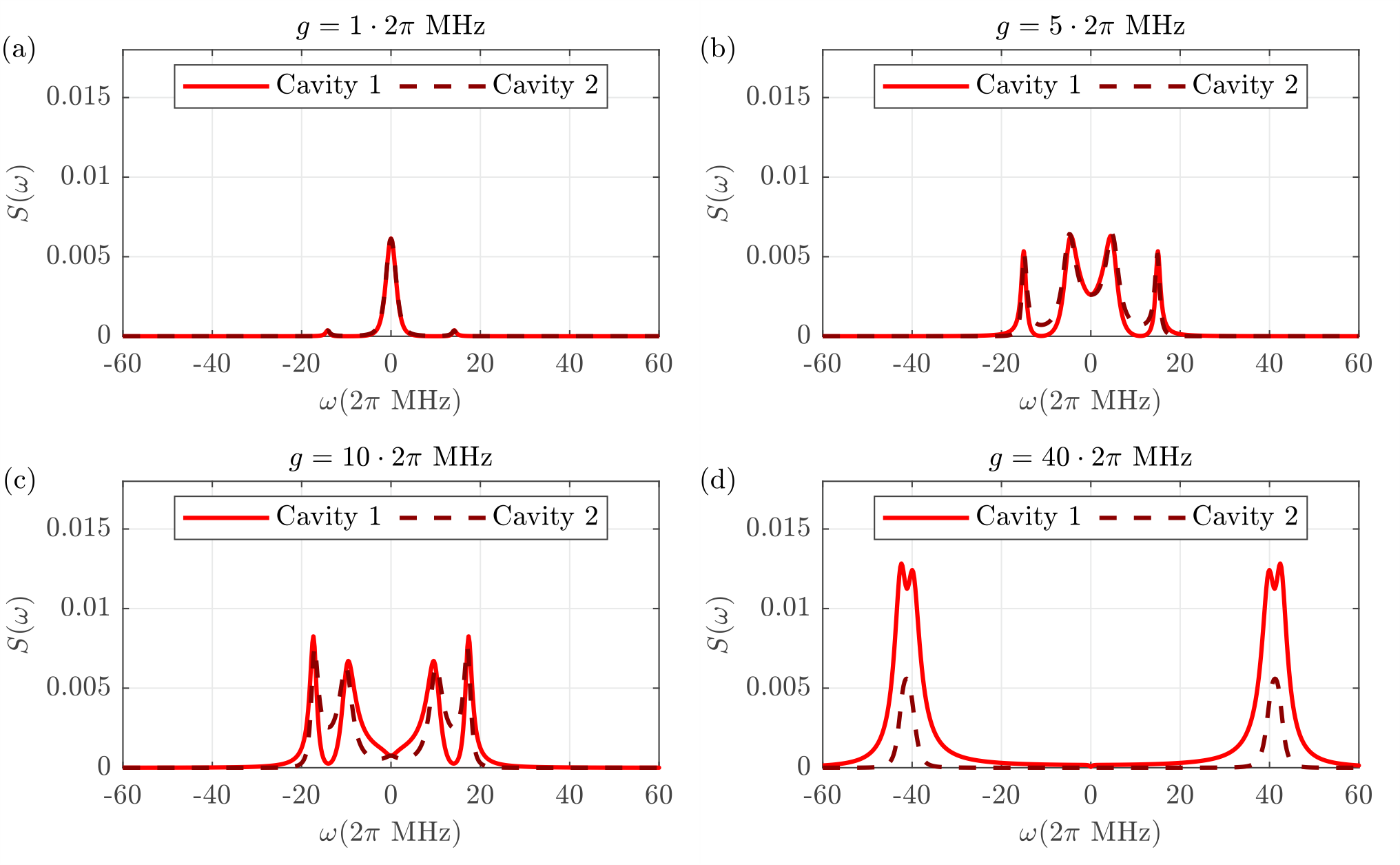}
	\vglue -.5cm
	\caption{Spectra of spontaneous emission from Cavities 1 and 2 for a range of atom-cavity coupling strengths. $[\kappa,\kappa_b,\gamma,v]=[1,0.01,5.2,10]$. Atom 1 is initially excited. }
		\label{fig:4Spectra}
\end{figure*}
\end{widetext}

The first two terms of this expression should not be surprising -- they tell us that the excitation of both the quasi-bright state and  fiber-dark modes contribute to its behaviour. Indeed, in the limit $\Delta_{S\pm} \rightarrow 0$, this expression collapses to a simple sum of the two, as expected from (\ref{eqns:opticalintermsofnormal}). There is, however, a contribution from the quasi-cavity-dark mode, on the order of $(\Delta_{S+} - \Delta_{S-})$. It can be difficult to quantify this contribution in the time domain -- we thus resort to the spectral behaviour to investigate this further. 

\section{Spectrum of Spontaneous Emission} \label{sec:SpectrumSpontaneousEmission}
The spectrum of spontaneous emission, written as follows for one of the atoms, can be found by taking the double integral of the appropriate two-time correlation function \cite{Carmichael1997,Carmichael2007}

\begin{equation}
T(\omega) = \frac{\gamma}{2\pi}\int_{0}^{\infty} dt \int_{0}^{\infty} dt' e^{-i\omega (t-t')} \left< \sigma_i^+(t)\sigma_i^-(t')\right>,
\label{eq:SpontaneousEmission}
\end{equation}

Using the quantum regression theorem, it can be shown that the spectrum is given by the squared modulus of the Laplace transform $\tilde{\xi}_i(s)$ of the probability amplitude at $s=-i\omega$:

\begin{equation}
	T(\omega) = \frac{\gamma}{2\pi} \abs{\tilde{\xi}_i(-i\omega)}^2.
	\label{eq:SponEmissSpectrumFormula}
\end{equation}

The spectrum of emission from the fiber or cavity outputs are given similarly, with the appropriate probability amplitudes and decay rates.

Figure \ref{fig:4Spectra} demonstrates the spectrum of spontaneous emission from the two cavities, calculated in a similar manner:
\begin{equation}
    S(\omega) = \frac{\kappa}{\pi}\abs{\tilde{\alpha}_i(-i\omega)}^2,\quad (i = 1,2),
\end{equation}

These cavity spectra are calculated for a variety of different atomic coupling strengths. We see that at very low values of $g$, the two fiber-dark modes are indistinguishable, forming a single peak centered at $\omega=0$. There is a small contribution from the two bright states, centered at approximately the normal mode splitting $\pm\zeta$. As $g$ is increased, the fiber-dark modes become more resolvable, until greatly exceeding $v$, where the contributions from the bright states and fiber-dark modes are almost indistinguishable.

The features in the spontaneous emission spectrum can be used to explain the behaviour observed in the cavity occupation. For example, Figure \ref{fig:4Spectra} (a) confirms that the small oscillations in the cavity occupation in the fiber-dominated coupling limit in Figure \ref{fig:AtomDomOscillations} (a) are due to the small excitation of the bright states. Equivalently, in the atom-dominated case in Figure \ref{fig:4Spectra} (d), the origin of the oscillations in the cavity occupation is clearly due to an equal excitation of both the fiber-dark modes and bright states.

\subsection{Contribution of interference}

We demonstrated in (\ref{eq:Eqnswithfivecoeffs}) that the original mode probability amplitudes propagate with at most five eigenvalues of the non-Hermitian Hamiltonian $\mathcal{H}$. Taking the Laplace transform of the coefficients yields

\begin{equation}
\tilde{c}_i(-i\omega) = \sum_{j=1}^{5} \chi_{ij} L(\omega,\lambda_j),
\end{equation}
where $L(\omega,\lambda_j)$ is the spectral function
\begin{equation}
L(\omega,\lambda_j)=\frac{1}{\eta_j - i(\omega -\delta_j)},
\end{equation}
where the eigenvalues $(\lambda_j)$ are split into real $(\eta_i)$ and imaginary $(\delta_j)$ parts,
\begin{equation}
\lambda_j = \eta_i + i\delta_j.
\end{equation}
Each spectral function $L(\omega,\lambda_j)$ and coefficient $\chi_{ij}$ corresponds to the relative excitation and contribution from a quasi mode to the optical output. It is now clear from (\ref{eq:SponEmissSpectrumFormula}) that the optical spectra can be expressed as the sum of at most five Lorentzian functions, $\abs{L(\omega,\lambda_j)}^2$ and ten interference terms, $W_{jk}(\lambda_j,\lambda_k)$, such that
\begin{equation}
\abs{\tilde{c}_i(-i\omega)}^2 = \sum_{j=1}^{5}\abs{\chi_{ij}}^2 \abs{L(\omega,\lambda_j)}^2  + \sum_{j<k} W_{jk}(\omega,\lambda_j,\lambda_k),
\label{eq:Expressionforspectrafull}
\end{equation}
where we quantify the `interference' between two quasi modes as
\begin{equation}
W_{jk}(\omega,\lambda_j,\lambda_k) = \chi_{ij}\chi_{ik}^* L(\omega,\lambda_j)L^*(\omega,\lambda_k) + \textrm{c.c.}
\end{equation}
These interference functions are not strictly positive or negative. However, it is informative to calculate their integral, to find their effective net contribution to the cavity output:

\begin{equation}
\int_{-\infty}^{\infty} W_{jk}(\omega,\lambda_j,\lambda_k) d\omega = \frac{2\pi \chi_{ij}\chi_{ik}^*}{(\delta_i - \delta_j)i + (\eta_1 + \eta_2)} +\textrm{c.c.}
	\label{eqn:InterferenceQuantity}
\end{equation}

\begin{figure}
	\includegraphics[width =\columnwidth]{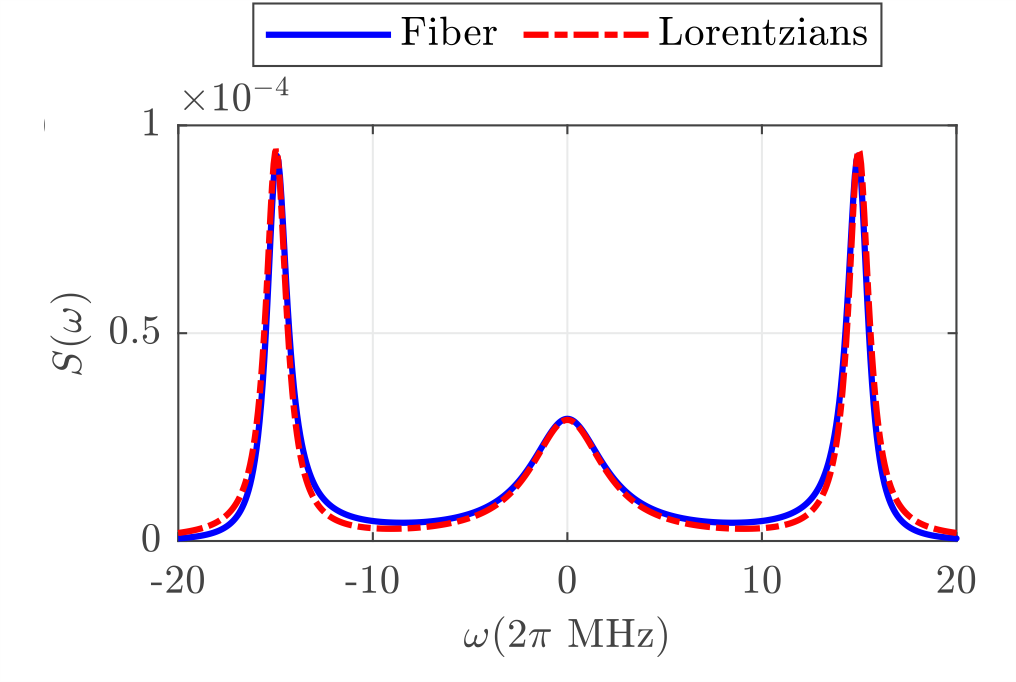}
		\vglue -.5cm
	\caption{Emission spectrum from the fiber. The red curve indicates the approximation as the sum of three Lorentzian functions. $[\kappa,\kappa_b,g,v]=[1,0.01,5,10]$.}
	\label{fig:EmissionFiberandApprox}
\end{figure}

From (\ref{eqn:InterferenceQuantity}), it should be clear that, regardless of the coefficients $\chi_{ij},\chi^*_{jk}$, the interference functions have minimal contribution to the cavity spectra when $\delta_i - \delta_j \gg 0$. In such cases the modes are well separated in frequency, and there is no significant overlap of the Lorentzian spectral components. The spectrum is well approximated by the sum of these Lorentzian functions, and the interference terms can be disregarded. For example in Figure \ref{fig:EmissionFiberandApprox} in the fiber output,  $S(\omega) = \frac{\kappa_b}{\pi}\abs{\tilde{\beta}(-i\omega)}^2$, the quasi-fiber-dark modes are not excited and the normal mode splitting $\zeta$ is large, and thus interference effects are minimal.

\begin{figure}
	\includegraphics[width =\columnwidth]{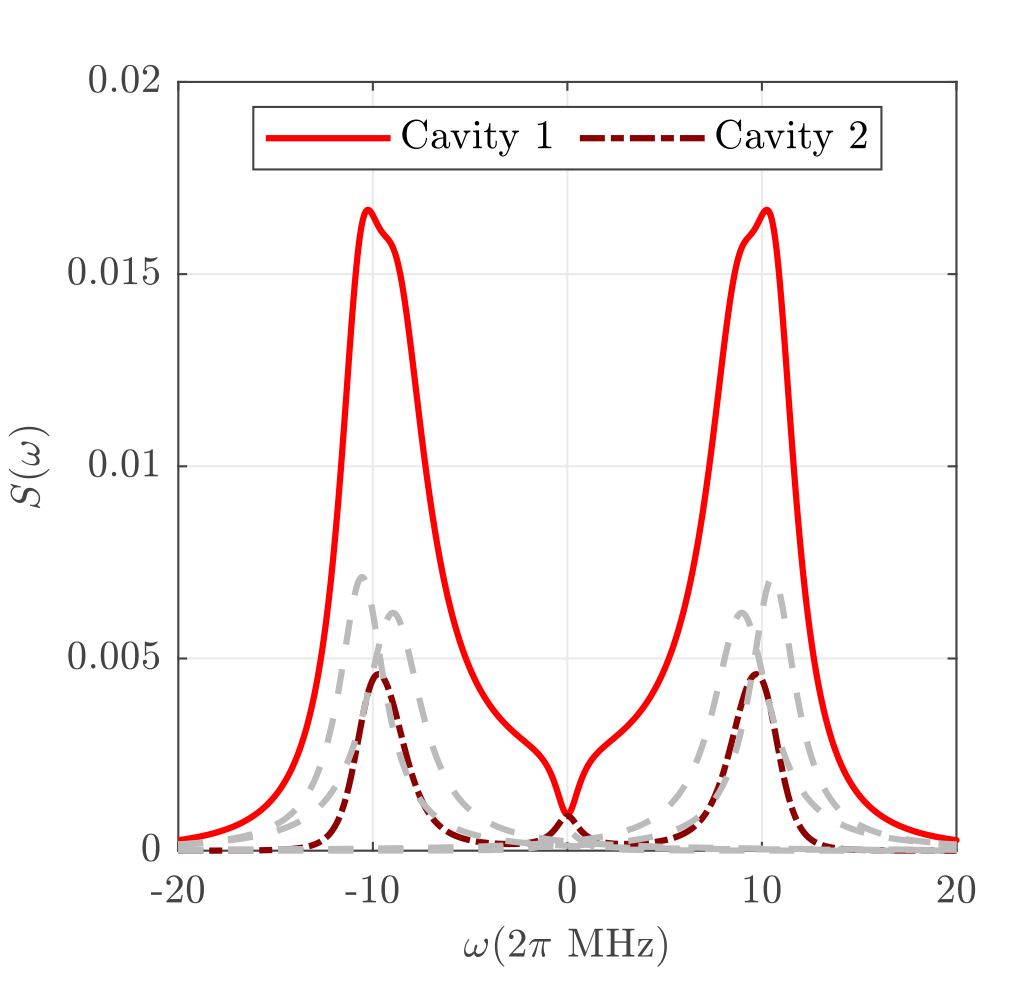}
	\vglue -.5cm
		\caption{Emission spectra from Cavity 1 and Cavity 2. The grey curves are the five excited Lorentzians, corresponding to the excitation of each quasi-normal mode. $[\kappa,\kappa_b,g,v]=[1,0.01,7,4]$.}
	\label{fig:BothCavities}
\end{figure}

However, when the atom-dominated coupling regime is approached and the normal mode splitting $\zeta$ does not greatly exceed $g$, we find that the interference effects are significant. Consider Figure \ref{fig:BothCavities}, where the emission spectrum from Cavity 1 has a much greater spectral intensity at all frequencies except on resonance, where there is a significant drop in intensity in Cavity 1 relative to Cavity 2. Such an effect cannot be explained by the excitation of the quasi cavity-dark mode alone, as there is a net negative contribution to one of the spectral outputs.

To quantify these interference effects, we must simply look at the coefficients $\chi_{ij}$, which are obtained from (\ref{eqns:QuasiBrightStateTimePerturbed}), (\ref{eqn:ApproxtoCavitiesTime}) and (\ref{eqn:AnalyticalFiberDark}) :

\begin{equation}
\begin{aligned}
	\chi_{C_1,BS\pm} &\approx \pm\frac{1}{2\zeta} \left[\left(\frac{g}{2}- v\Delta_{S\pm}\right)\right],\\
	\chi_{C_1,FD\pm} &= \pm \frac{g}{4p},\\
	\chi_{C_1,CD} &\approx \frac{(\Delta_{S+} -\Delta_{S-})}{2\zeta}\left[\frac{g}{2}(\Delta_{S+} + \Delta_{S-}) + v\right],
\end{aligned}
\label{eqn:CavContributions}
\end{equation}
and

\begin{equation}
\begin{aligned}
	\chi_{C_1,BS\pm} & =\chi_{C_2,BS\pm} \\
	\chi_{C_1,FD\pm} &= -\chi_{C_2,FD\pm},\\
	\chi_{C_1,CD} &= 	\chi_{C_2,CD}.
\end{aligned}
\end{equation}

\begin{figure}
	\includegraphics[width =\columnwidth]{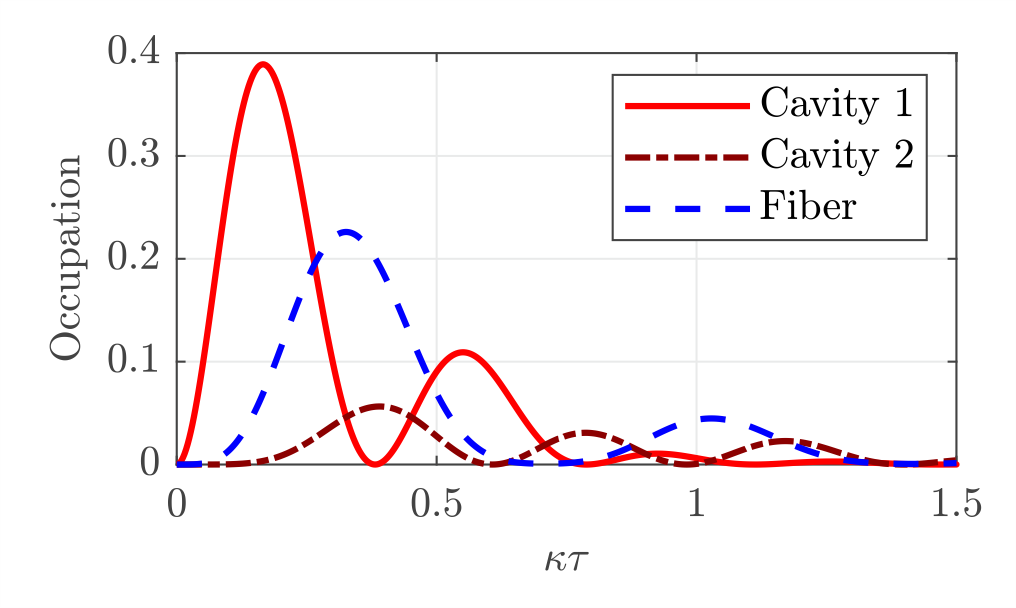}
	\vglue -.5cm
		\caption{Relative excitations of the optical modes and fiber mode, with the same parameters as in Figure \ref{fig:BothCavities}. Note that in this parameter regime, the occupation of the fiber is significant.}
	\label{fig:timedependentbothsimilar}
\end{figure}

This translates to a rather simple result: each cavity will share the same five fundamental Lorentzians, with the same interference effects between these three symmetric modes. Figure \ref{fig:BothCavities} displays this concept, with the five Lorentizans of the system given in grey. It is clear from (\ref{eqn:CavContributions}) that the Lorentzian contribution from the quasi cavity-dark mode should be minimal, and reduces as the normal mode splitting increases -- there is no visible contribution in Figure \ref{fig:BothCavities}. Once again, the time evolution of the optical occupations in Figure \ref{fig:timedependentbothsimilar} shows how the excitation travels from cavity 1 to cavity 2 via the fiber mode.

\begin{figure}
	\centering
	\includegraphics[width =\columnwidth]{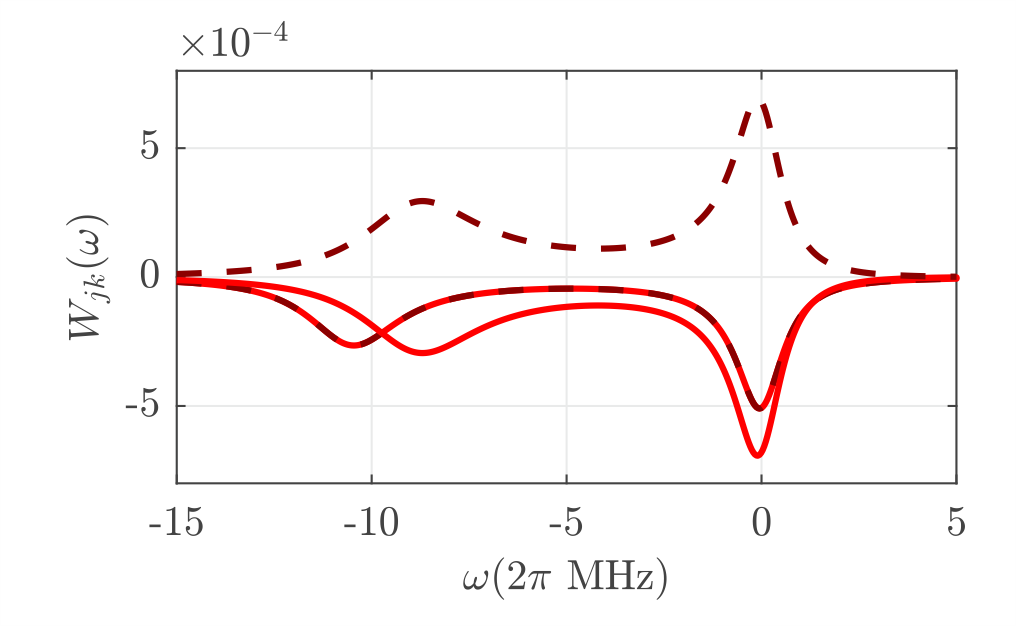}
	\vglue -.5cm
		\caption{Interference effects $W_{jk}(\omega)$ between the quasi cavity-dark mode and the qBS$_-$ and qFD$_-$ modes for Cavity 1 (solid) and Cavity 2 (dashed).  $[\kappa,\kappa_b,g,v]=[1,0.01,9,4]$.}
	\label{fig:CDInterference}
\end{figure}

In contrast, the interference between a symmetric and an anti-symmetric mode will have an equal and opposite contribution between Cavity 1 and 2. Consider Figure \ref{fig:CDInterference}, where four interference functions $W_{jk}$ are plotted -- the interactions between the quasi cavity-dark mode, and the quasi fiber-dark and bright states. As the cavity-dark and bright states are symmetric, their respective contributions to Cavity 1 and 2 are the same -- however, the interaction between the cavity-dark and fiber-dark quasi modes yields a positive contribution to Cavity 2, but a negative contribution to Cavity 1. The sum of the relative interferences yields a small increase in spectral intensity on resonance for Cavity 2, and a large decrease for Cavity 1. The same explanation can be used for the differences in the spectral intensity at $\omega \approx \zeta$, in terms of interferences between the fiber-dark quasi modes and quasi bright states.

Unsurprisingly, this interference effect can be entirely eliminated by setting $\kappa_b =\gamma/2$ $(\Gamma_{SD}=0)$ -- this will suppress the cavity-dark quasi mode, and no unusual effects would be noticed on resonance. In terms of (\ref{eqns:MotionofNormalModesOriginal}), this amounts to decoupling the cavity-dark mode from the bright states, and thus preventing it from being observable from the cavity outputs.

\section{Conclusion}
We have discussed the dynamical behaviour of the optical and normal modes of an all-fiber cavity-QED system in the single excitation limit, applying both an analytic and perturbative approach. Using the spectrum of spontaneous emission, we have quantified the excitation of certain normal modes in different parameter regimes, and related these to the oscillations observed in the occupation of the cavity modes. We also provide a means of approximating the spectrum of spontaneous emission through first order perturbation theory, which allows us to make accurate predictions of the system's behaviour and easily identify the origin of unusual interference effects.

A natural candidate for future research is to consider the behaviour of the system when both atoms begin in the excited state; by taking a similar trajectory approach, one could explore the time-dependent occupation of the cavities, and  calculate correlations between the different decay channels of the system in different parameter regimes. Additionally, one could also consider the influence of a finite time delay for a system with a sufficiently long fiber; this would introduce non-Markovian dynamics and, for the case of a single excitation, would result in a set of delayed differential equations for the probability amplitudes \cite{EntanglementDelay,WhalenThesis,BenoitTransfer,Non-Markov2019}.

\section*{Acknowledgments}
The authors thank Takao Aoki and Donald White for interesting and helpful discussions. 

\appendix
\section{Further Perturbative Approach}
In the case that $v < g/2$ and $\Gamma_{SD} > \Gamma_{D}$, it is still appropriate to include $\Gamma_D$ in the unperturbed Hamiltonian $\mathcal{H}_0$. Indeed, if we consider $\Gamma_D$ in the perturbation, we find the equations to be of the same form as (\ref{eqns:pertrightstates}), albeit with
\begin{equation}
\begin{aligned}
\Delta_{S\pm} = \frac{\Gamma_{SD}}{-\frac{\Gamma_{S+}}{2} \mp i\zeta},
\end{aligned}
\end{equation}
and a first order shift to the quasi cavity-dark mode eigenvalue
\begin{equation}
\Delta\lambda_D = -\Gamma_D.
\end{equation}
This in fact \textit{decreases} the accuracy of the perturbation, as there is no initial `guess' of the true eigenvalue.

While the given perturbative approach in Sec. (\ref{sec:Pertapproach}) is valid, we can make several adjustments to improve the accuracy of the approach, and broaden the parameter space in which this approximation can be made. Perhaps the simplest modification is to calculate the second order shifts in the eigenvalues: The most obvious modification to the perturbation $P$ is to include the interaction $\Gamma_{S-}/2$ between the two bright states. This modifies the quasi bright states to be of the form

\begin{equation}
		\ket{QBS_{\pm}}_1 = \ket{BS_\pm} \mp \frac{i\Gamma_{S-}}{4\zeta} \ket{BS_\mp} + \Delta_{S\pm}\ket{CD}.
\end{equation}

Additionally, it is possible to further increase the accuracy of the perturbation by calculating the second order shifts to the eigenvalues.

\begin{equation}
\begin{aligned}
\Delta^{(2)}\lambda_{CD} &= \bra{CD}P\ket{QCD}_1=-\Gamma_{SD}(\Delta_{S+} + \Delta_{S-}),\\
\Delta^{(2)}\lambda_{BS_\pm} &= \bra{BS_\pm}P\ket{BS_\pm}_1= \pm \frac{i \Gamma_{S-}^2}{8\zeta} + \Gamma_{SD}\Delta_{S\pm}.
\end{aligned}
\end{equation}


\

\bibliographystyle{apsrev4-1}

\end{document}